\documentclass[12pt]{article}
\usepackage{graphicx}
\newcommand{\sss}{\scriptscriptstyle}

\newcommand {\be}{\begin{equation}} 
\newcommand{\ee}{\end{equation}}    

\def\dds1{\frac{\partial}{\partial s_1}}

\def\vtj{v_{{\sss T}j}}

\def\d{d\kern-0.8 ex\vrule height 1.3 ex depth-1.24 ex width 0.7 ex
\kern 0.15 ex}
\def\D{D\kern-1.7 ex\vrule height .87 ex depth-0.8 ex width 0.7 ex
\kern 0.95 ex}

\begin{document}
\baselineskip 20 pt

\begin{center}

\Large{\bf Current-less solar wind driven dust acoustic instability in cometary plasma }

\end{center}

\begin{center}

 J. Vranjes

{\em Belgian Institute for Space Aeronomy,  Ringlaan 3,
1180 Brussels, Belgium}

\end{center}

\vspace{2cm}

{\bf Abstract:}
A quantitative analysis is presented of the dust acoustic wave instability driven by the solar and stellar winds. This is a current-less kinetic instability which develops in permeating plasmas, i.e..,   when one quasi-neutral electron-ion wind plasma in its propagation penetrates  through another quasi-neutral plasma which contains dust, electrons and ions.

\vspace{1cm}


\vfill

\pagebreak

During the previous passage of the comet Halley, waves and  oscillations in the cometary plasma, some of them with very low frequencies  in the range of 10 Hz, were clearly observed.$^{1, 2}$
 Such a cometary plasma itself is a rather complex system which includes charged grains   and  various ion species$^{3}$,  
  depending on the distance from the coma. Those include ions like  $H^+$, $H_2^+$, $C^+$, $O^+$, $OH^+$, $H_2O^+$, etc, with pick-up ions $H^+$ from the solar wind being most abundant in the outer regions. The presence of the solar wind plasma, which  penetrates  the cometary cloud, adds to  the complexity and makes the whole system potentially unstable due to the free energy stored in the streaming solar wind species.

In our recent work$^4$ 
 a current-less instability of the ion acoustic (IA) perturbations was demonstrated in the case of one quasi-neutral electron-ion plasma propagating through another static quasi-neutral (target) plasma. The threshold
velocity of the propagating plasma was shown to be in principle  even  below the ion acoustic (IA) speed of the static plasma, the latter (IA speed) being known to be  the threshold speed for the classic case of the electron current driven IA wave instability. Hence, the acoustic instability in such permeating plasmas may be much more effective than in classic examples with current-carrying plasmas,  and it may be frequently expected in space and astrophysical plasmas where such configurations are rather common. Being current-less, it also removes the problem of the self-induced magnetic field associated with the background electron current in the classic instability.  In the present work this will be investigated in the case of solar (stellar) winds propagating through the surrounding interplanetary (interstellar) plasmas that also frequently include dust, which may be of cometary or any other origin.$^5$ 
This implies acoustic perturbations which involve the dust dynamics, i.e., the dust acoustic mode that has continuously  been in the focus of researchers$^{6-14}$ 
ever since its prediction$^{15}$ 
and experimental verification.$^{16}$. 

One may start from the linearized Vlasov-Boltzmann kinetic equation for the perturbed distribution function
\[
\frac{\partial f_{j1}}{\partial t} + \vec v \frac{\partial
f_{j1}}{\partial\vec r} + \frac{q_j }{m_j} \vec
E_1 \frac{\partial f_{j0}}{\partial \vec v} = 0,
\]
where  the plasma distribution function for the species $j$ is
\be
f_{j0}=\frac{n_{j0}}{(2\pi)^{3/2}\vtj^3} \exp\left\{-\frac{1}{2
\vtj^2}\left[v_x^2 + v_y^2 + (v_z- v_{j0})^2\right]\right\},
\label{e1} \ee
Here, $n_{j0}=const$, and $\vtj^2= \kappa T_j/m_j$, and we assume  longitudinal perturbations $\sim \exp(-i
\omega + i k z)$ propagating in the direction of  the interplanetary magnetic field vector. In such a geometry the effects of the magnetic field will not affect the perturbations.

The perturbed number density may be calculated from
\[
n_{j1}=\int f_{j1} d^3 \vec v.
\]
For the general species $j$ this yields
\be
\frac{n_{j1}}{n_{j0}}=-\frac{q_j \phi_1}{\kappa T_j}\left[1- {\cal Z}(\alpha_j)\right]
\label{e2}
\ee
For non-streaming species $\alpha_j=\omega/(k \vtj)$, and
\be
 {\cal Z}(\alpha_j) =\frac{\alpha_j}{(2 \pi)^{1/2}} \int d\xi \exp(-\xi^2/2)/(\alpha_j-\xi).\label{e3}
 \ee
 The integration in (\ref{e3}) is along the Landau contour, and $\xi=v_z/\vtj$.

 For the streaming species the derivation is similar and Eq.~(\ref{e2}) is obtained, but instead of $\alpha_j$ and $\xi$ now we have $\beta_j=(\omega- k v_0)/(k \vtj)$ and   $\zeta=(v_z-v_0)/\vtj$. The quasi-neutrality in the perturbed state
 \be
 n_{wi1}+  n_{ci1}=n_{we1}+ n_{ce1} + Z_d n_{d1}\label{e4}
 \ee
 will directly yield the dispersion equation. Here, and further in the text, for the cometary and wind  plasma the indices $c$ and $w$ are used.

In Eq.~(\ref{e2}) for dust the following expansion  (to be supported  by numbers  in the text below) will be used
 \be
  {\cal Z}(\alpha_d)\simeq 1+ \frac{1}{\alpha_d^2} + \frac{3}{\alpha_d^4} + \ldots - i \left(\frac{\pi}{2}\right)^{1/2} \alpha_d \exp(-\alpha_d^2/2).
  \label{dex}
  \ee
 This is valid if $|\alpha_d|\gg 1$ and $|Re(\alpha)_d|\gg |Im(\alpha_d)|$.
For the two electron populations we shall use  ${\cal Z}(\alpha_{e})\simeq -i (\pi/2)^{1/2} \alpha_{e}$, ${\cal Z}(\beta_e)\simeq -i (\pi/2)^{1/2} \beta_e$, that is valid for
\[
|\alpha_{e}|\equiv \frac{|\omega|}{k v_{{\sss T}ce}} \ll 1, \quad |\beta_{e}|\equiv \frac{|\omega- k v_0|}{k v_{{\sss T}we}}\ll 1.
 \]
 The same will be used for the cometary ions, $|\alpha_i|\equiv |\omega|/(k v_{{\sss T}ci}) \ll 1$.  As for the wind ions, the following two interesting cases can be studied.


 The parameter $\beta_{wi}=(\omega- k v_0)/(k v_{\sss{T}wi})$ contains two terms, where for the first one we expect that  $\omega/(k v_{\sss{T}wi})\ll 1$, while for the second one in this case we assume $ v_0/v_{\sss{T}wi}\gg 1$. Hence, the expansion similar to (\ref{dex}) should be used.
The dispersion equation in general form reads
\[
\frac{z_{ci}^2 n_{ci0}}{T_{ci}} \left[1+i\left(\frac{\pi}{2}\right)^{1/2} \frac{\omega}{k v_{{\sss T}ci}}\right]
+ \frac{n_{we0}}{T_{we}} \left[1+i\left(\frac{\pi}{2}\right)^{1/2} \frac{\omega- k v_0}{k v_{{\sss T}we}}\right]
+\frac{n_{ce0}}{T_{ce}} \left[1+i\left(\frac{\pi}{2}\right)^{1/2} \frac{\omega}{k v_{{\sss T}ce}}\right]
\]
\[
-\frac{z_{wi}^2 n_{wi0}}{T_{wi}}\left\{\frac{k^2 v_{{\sss T}wi}^2}{(\omega-k v_0)^2} - i\left(\frac{\pi}{2}\right)^{1/2} \frac{\omega- k v_0}{k v_{{\sss T}wi}}\exp\left[\frac{(\omega- k v_0)^2}{2 k^2 v_{{\sss T}wi}^2}\right]\right\}
\]
\be
-\frac{z_d^2 n_{d0}}{T_d}\left\{\frac{k^2 v_{{\sss T}d}^2}{\omega^2} - i\left(\frac{\pi}{2}\right)^{1/2} \frac{\omega}{k v_{{\sss T}d}}\exp\left[\frac{\omega^2}{2 k^2 v_{{\sss T}d}^2}\right]\right\}\equiv Re[\Delta(k, \omega)] + i Im[\Delta(k, \omega)]=0.\label{d1}
\ee
In application to the solar wind interaction with cometary dusty plasma, one can make a few  simplifications in Eq.~(\ref{d1}).
 The cometary plasma parameters  are used from Ref. 17. 
 These imply the following: $T_{ce}=1.16 \cdot 10^5$ K ($=10$ eV), $T_{ci}=2.32 \cdot 10^4$ K ($=2$ eV), $T_d=1.16\cdot 10^2$ K ($=0.01$ eV), $n_{ci0}=10^7$ m$^{-3}$, $n_{d0}=10$ m$^{-3}$, $Z_d=800$, $m_d=1.13 \cdot 10^{-20}$ kg, {\bf and the mean dust grain radius $a_d=3\cdot 10^{-8}$ m.} Similar data about the dust mass may be found also in Ref. 18 
  dealing with the latest passage of the comet Halley.
As for the electrons and ions from the solar wind, we use the following parameters: $T_{ew}=T_{iw}=1.5 \cdot 10^5$ K ($\simeq 13$ eV), $n_{wi0}= n_{we0}=5 \cdot 10^6$ m$^{-3}$. The solar wind speed is adopted to be $v_{we0}=v_{wi0}=v_0=5 \cdot 10^5$ m/s. Singly charged ions are assumed in both systems, and the grains are negatively charged so that
\be
n_{ce0}+ Z_d n_{d0}=n_{ci0}. \label{qn1}
\ee
The corresponding thermal speeds that will be used  are $v_{{\sss T}wi}=3.52\cdot 10^4$ m/s,  $v_{{\sss T}ci}=1.38\cdot 10^4$ m/s,
$v_{{\sss T}we}=1.5\cdot 10^6$ m/s, $v_{{\sss T}ce}=1.3\cdot 10^6$ m/s, $v_{{\sss T}d}=0.38$ m/s.

Note that for the interplanetary magnetic field of $B_0=5 \cdot 10^{-9}$ T,  the corresponding  gyro-radii for the two populations of protons and electrons are $\rho_{wi}=74$ km,  $\rho_{ci}=29$ km, $\rho_{we}=1.7$ km, $\rho_{ce}=1.5$ km.

With all this it is seen that, first, the contribution of the real term from the ion wind part (the terms with the $wi$-index) is negligible. Second, the ratio of the ion and electron imaginary terms is $z_{wi}^2(T_{we}/T_{wi})^{3/2} (m_{wi}/m_e)^{1/2} \exp[-v_0^2/(2 v_{{\sss T}wi}^2)]$ that is completely negligible. So the total contribution of the wind ions can be omitted. Third, the imaginary term from the cometary electrons is negligible as compared to the imaginary term from the cometary ions; the ratio of the two is proportional to  $ (n_{ce0}/n_{ci0}) (T_{ci}/T_{ce})^{3/2} (m_e/m_{ci})^{1/2}\ll 1$.

As a result, the real part of (\ref{d1}) yields the frequency of the dust acoustic mode
\be
\omega_r^2\simeq \frac{k^2 Z_d^2 n_{d0}}{n_{ce0}} \frac{\kappa T_{ce}}{m_d}  \frac{1}{1+  \frac{\displaystyle{n_{we0}}}{\displaystyle{n_{ce0}}}
 \frac{\displaystyle{T_{ce}}}{\displaystyle{T_{we}}} + \frac{\displaystyle{z_{ci}^2 n_{ci0} T_{ce}}} {\displaystyle{n_{ce0} T_{ci}}}   }.
\label{e5}
\ee
The growth rate  $\gamma \simeq - Im \Delta(k,
\omega_r)/[\partial (Re\Delta)/\partial \omega]_{\omega\simeq
\omega_r}$ is
\be
\gamma=\Upsilon \left\{ v_0 - \frac{\omega_r}{k} \left[ 1+ \frac{z_{ci}^2 n_{ci0}}{n_{we0}} \left(\frac{T_{we}}{T_{ci}}\right)^{3/2} \left(\frac{m_{ci}}{m_e}\right)^{1/2} \right]\right\}, \label{gam}
\ee
where
\[
\Upsilon=\left(\frac{\pi}{8}\right)^{1/2} \frac{n_{we0}}{Z_d^2 n_{d0}} \frac{m_d m_e^{1/2}}{(\kappa T_{we})^{3/2}} \frac{\omega_r^3}{k^2}.
\]
Hence, the instability sets in if
\be
v_0> \frac{\omega_r}{k} (1+ a), \quad a= \frac{z_{ci}^2 n_{ci0}}{n_{we0}} \left(\frac{T_{we}}{T_{ci}}\right)^{3/2} \left(\frac{m_{ci}}{m_e}\right)^{1/2}. \label{c1}
\ee
For the parameters given above the threshold velocity $v_{th}$ for the instability is in fact very low. Taking  $Z_d=800$, from (\ref{c1})  we have $v_{th}=5.3$ km/s only.  Therefore, {\bf within the present model} the wind-driven dust acoustic oscillations are always growing.   In Fig.~1 the growth rate is given in Hz, in terms of the wave number and the dust charge number density.

The wave frequency contour plot has the same shape, with the maximum frequency around 2 Hz. In the case of the  fast solar wind $v_0=8\cdot 10^5$ m/s and  for $Z_d=800$,  the ratio $\gamma/\omega_r$  is increased by about 60 percent.

In such a multi-component system, the Debye length is determined by the coolest species (the dust in the present case), and it turns out to be of the meter size. Hence, the shortest wavelengths of interest here should be above that scale and consequently the largest wave frequency that should be attributed to the DA mode is expected to be of the order of a few Hz.  In view of large gyro-radii given previously in the text, the magnetic field effect is clearly completely negligible even for the lightest species.

In the three-component cometary plasma  the DA phase speed is
\[
c_{\sss{DA}}=\left[\frac{Z_d^2 n_{d0} \kappa T_{ci} T_{ce}}{m_d (n_{ci0} T_{ce} + n_{ce0} T_{ci})}\right]^{1/2}.
\]
For the same parameters as above we have $c_{\sss{DA}}=3.9$ m/s. In the presence of the wind, the actual DA  phase speed $v_{ph}=\omega_r/k=3.8$ m/s. Observe that the threshold velocity is $v_{th}=1.4 v_{ph}$.

In the case
\be
|\omega- k v_0|/(k v_{\sss{T}wi})\ll 1
\label{exp2}
\ee
the expansion for the wind ions is different and it yields
\be
\frac{n_{wi1}}{n_{wi0}}=-\frac{e z_{wi}  \phi_1}{\kappa T_{wi}}\left[1 + i\left(\frac{\pi}{2}\right)^{1/2} \frac{\omega- k v_0}{k v_{{\sss T}wi}}\right].
\label{e22}
\ee
A procedure similar as above yields the frequency
\be
\omega_r^2\simeq \frac{k^2 Z_d^2 n_{d0}}{n_{ce0}} \frac{\kappa T_{ce}}{m_d}  \frac{1}{1+  \frac{\displaystyle{n_{we0}}}{\displaystyle{n_{ce0}}}
 \frac{\displaystyle{T_{ce}}}{\displaystyle{T_{we}}} + \frac{\displaystyle{z_{ci}^2 n_{ci0} T_{ce}}} {\displaystyle{n_{ce0} T_{ci}}} + \frac{\displaystyle{z_{wi}^2 n_{wi0} T_{ce}}} {\displaystyle{n_{ce0} T_{wi}}}  }.
\label{e55}
\ee
The corresponding growth-rate
\be
\gamma=\Upsilon \left[ v_0(1+ b) - \frac{\omega_r}{k} (1+ a + b)\right], \quad b= \left(\frac{T_{we}}{T_{wi}}\right)^{3/2} \left(\frac{m_{wi}}{m_e}\right)^{1/2} \frac{z_{wi}^2 n_{wi0}}{n_{we0}}, \label{g2}
\ee
yields the instability condition which in this case reads
\be
v_0> \frac{\omega_r}{k} \left(1+ \frac{a}{1+ b}\right). \label{c2}
\ee
For the solar wind data the parameter $b\simeq 43$, so the critical velocity (\ref{c2}) is reduced.

To conclude, the demonstrated current-less instability is expected to be very effective in the environment of cometary tails, and in interstellar dusty plasma penetrated by the stelar winds in general. The threshold velocity is very low, practically equal to the phase speed of the DA wave propagating in such a two-plasma system [this particularly in the limit (\ref{exp2})], and the restoring pressure appears mainly due to the contribution of the cometary electrons.  The model presented here is based on the assumption of a  fixed charge on dust grains. However, in realistic situations inelastic collisions  (i.e., those related to  charging of dust grain and absorption of plasma particles) may play an important role and yield considerably different results. These effects  have been studied in detail in Ref. 13 and in several references cited therein.$^{19}$ In the range of large wave-lengths the damping rate due to dust charging can become several orders of magnitude larger than the Landau damping on ions, while in the short wave-length domain the charging effects are almost absent. This short wave-length range is in fact of  interest in the present study because it provides a much more strongly growing mode, as can be seen from Fig.~1, and in this regime the charging phenomena should not considerably alter  the mode behavior.


 \pagebreak



\begin{figure}
\includegraphics[height=6cm, bb=30 0 750 550, clip=]{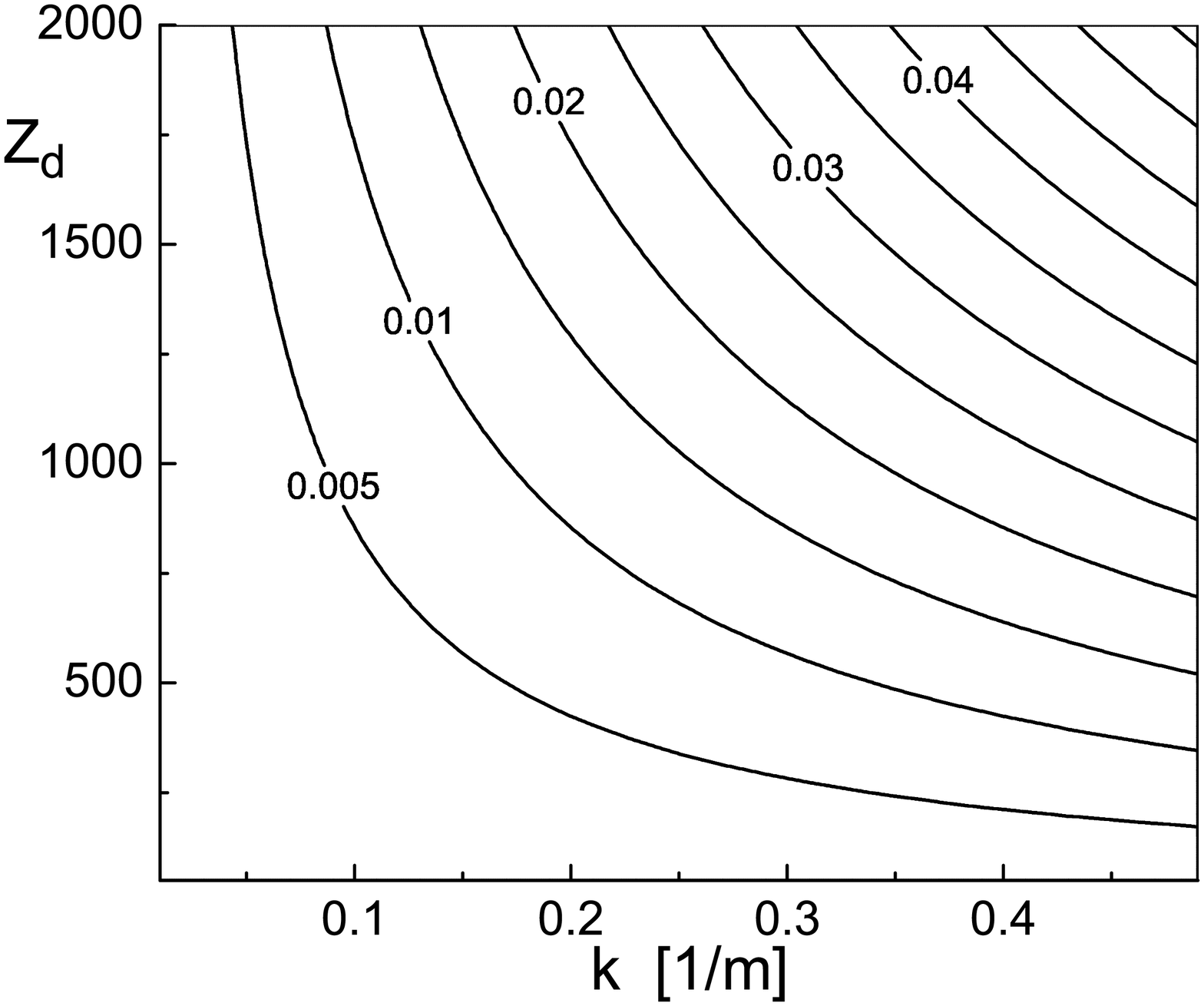}
\vspace*{-5mm} \caption{  The growth rate (in Hz) of the DA mode in a cometary dusty plasma, driven by the solar wind, in terms of the wave-number and the grain charge number.}\label{fig1}
\end{figure}

\end{document}